\documentclass[twocolumn]{aastex631}

\usepackage{amsmath}
\usepackage{enumitem}
\usepackage{multirow}
\usepackage{hyperref}

\begin{document}


\title{Investigating the Electron Capture Supernova Candidate AT 2019abn with JWST Spectroscopy}

\author[0000-0003-4725-4481]{Sam Rose} 
\correspondingauthor{Sam Rose}
\email{srose@caltech.edu}
\affiliation{Division of Physics, Mathematics, and Astronomy, California Institute of Technology, Pasadena, CA 91125, USA}

\author[0000-0003-0778-0321]{Ryan M.\ Lau}
\affil{NSF NOIRLab, 950 N. Cherry Ave., Tucson, AZ 85719, USA}

\author[0000-0001-5754-4007]{Jacob E.\ Jencson}
\affil{Caltech/IPAC, Mailcode 100-22, Pasadena, CA 91125, USA}

\author[0000-0002-5619-4938]{Mansi M. Kasliwal}
\affil{Division of Physics, Mathematics, and Astronomy, California Institute of Technology, Pasadena, CA 91125, USA}

\author[0000-0002-8989-0542]{Kishalay De}
\affil{MIT-Kavli Institute for Astrophysics and Space Research, 77 Massachusetts Ave., Cambridge, MA 02139, USA}

\author[0000-0001-5644-8830]{Michael E.\ Ressler}
\affil{Jet Propulsion Laboratory, California Institute of Technology, MS 169-327, 4800 Oak Grove Drive, Pasadena, CA 91109, USA}

\author[0000-0003-2238-1572]{Ori D.\ Fox}
\affil{Space Telescope Science Institute, 3700 San Martin Drive, Baltimore, MD 21218, USA}

\author[0000-0001-9315-8437]{Matthew  J.\ Hankins}
\affil{Arkansas Tech University, 215 West O Street, Russellville, AR 72801, USA}

\begin{abstract}

The James Webb Space Telescope (JWST) has opened up a new window to study highly reddened explosive transients. We present results from late-time (1421 days post-explosion) JWST follow-up spectroscopic observations with NIRSpec and MIRI LRS of the intermediate luminosity red transient (ILRT) AT 2019abn located in the nearby Messier 51 galaxy (8.6 Mpc). ILRTs represent a mysterious class of transients which exhibit peak luminosities between those of classical novae and supernovae and which are known to be highly dust obscured. Similar to the prototypical examples of this class of objects, NGC 300 2008-OT and SN 2008S, AT 2019abn has an extremely red and dusty progenitor detected only in pre-explosion \textit{Spitzer}/IRAC imaging at 3.6 and 4.5 $\mu$m and not in deep optical or near-infrared HST images. We find that late time observations of AT 2019abn from NEOWISE and JWST are consistent with the late time evolution of SN 2008S. In part because they are so obscured by dust, it is unknown what produces an ILRT with hypotheses ranging from high mass stellar merger events, non-terminal stellar outbursts, or terminal supernovae explosions through electron-capture in super-AGB stars. Our JWST observations show strong mid-IR Class C PAH features at 6.3 and 8.25 $\mu$m typical of carbon-rich post-AGB sources. These features suggest the dust around AT 2019abn, either pre-existing or newly formed in the ejecta, is composed of carbonaceous grains which are not typically observed around red supergiants. However, depending on the strength and temperature of hot bottom burning, SAGBs may be expected to exhibit a carbon-rich chemistry. Thus our JWST observations are consistent with AT 2019abn having an SAGB progenitor. 

\end{abstract}

\section{Introduction}
\label{sec:Introduction}

\begin{figure*}[t!]
    \centering
    \includegraphics[scale=0.7]{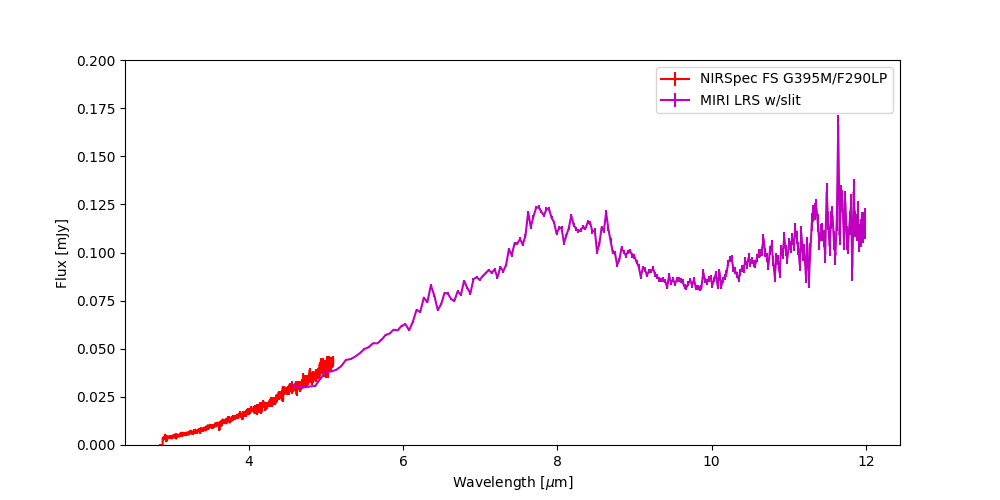}
    \caption{Combined NIRSpec and MIRI LRS spectra of AT 2019abn.}
    \label{fig:comb_spec}
\end{figure*}

Intermediate Luminosity Red Transients (ILRTs) represent a rarely detected and mysterious class of transients that may occur at a significant fraction ($\sim$ 5-10 $\%$) of the volumetric core collapse supernova rate \citep{Thompson:2009ApJ...705.1364T, Cai:2021A&A...654A.157C, Karambelkar:2023ApJ...948..137K}.
This class is characterized by luminosities between those of classical novae and supernovae \citep{Rau:2009PASP..121.1334R, Kasliwal:2012PASA...29..482K,Cai:2018MNRAS.480.3424C, Pastorello:2019NatAs...3..676P} and very red colors.
Like other classes of so-called ``gap transients", for example the SN impostors produced by Luminous Blue Variable (LBV) massive stellar eruptions \citep{Humphreys:1994PASP..106.1025H, Smith:2011MNRAS.415..773S, Smith:2014ARA&A..52..487S} and the Luminous Red Novae (LRNe) produced by stellar mergers \citep{Kulkarni:2007Natur.447..458K, Tylenda:2011A&A...528A.114T, Kochanek:2014MNRAS.443.1319K, Blagorodnova:2017ApJ...834..107B, Karambelkar:2023ApJ...948..137K}, more complete transient surveys in the last decade like the Zwicky Transient Facility \citep[ZTF;][]{Bellm:2019PASP..131a8002B, Graham:2019PASP..131g8001G, Dekany:2020PASP..132c8001D} and the Asteroid Terrestrial-impact Last Alert System \citep[ATLAS;][]{Tonry:2018PASP..130f4505T} have discovered many more of these low-luminosity extra-galactic transients.

While intermediate-luminosity red transient (ILRT) is sometimes used as a catch-all for a diverse set of events, in this work we use ILRT to specifically refer to a class of objects defined by two prototypical examples SN 2008S \citep{Botticella:2009MNRAS.398.1041B} and NGC 300 2008-OT \citep{Prieto:2008ApJ...681L...9P}. 
Both SN 2008S and NGC 300 2008-OT have lower mass, highly extincted progenitors which are only detected in infrared \textit{Spitzer} Space Telescope images and not in deep optical images obtained with the \textit{Hubble} Space Telescope.
The transients themselves are also very red suggesting high levels of extinction before, during, and after the outburst. 
The other ILRTs which closely resemble the prototypes include PTF10fqs, AT 2012jc, AT 2013la, AT 2013lb, AT 2017be, AT 2018aes, AT 2018hcj, AT 2019udc, AT 2019abn, AT 2019sfo, AT 2019ahd, AT 2020nqq, AT2021adlx, and AT 2022uqn \citep{Kasliwal:2011ApJ...730..134K, Boersma:2014ApJS..211....8B, Cai:2018MNRAS.480.3424C, Jencson:2019ApJ...880L..20J, Karambelkar:2023ApJ...948..137K, Deckers:2022TNSCR3036....1D}.
AT 2019krl, which is sometimes also called an ILRT in the literature, has a blue supergiant progenitor distinguishing it from those objects like SN 2008S and NGC 300 2008-OT \citep{Andrews:2021ApJ...917...63A} although the light curve and spectral properties are very similar.

ILRT light curves resemble the light curves of Type IIL or Type IIP core-collapse supernovae (CCSNe), albeit scaled down in luminosity.
Typically ILRTs rise to peak over a roughly 2-week period.
Following the peak there is a linear decline (Type IIL-like) or plateau (Type IIP-like) lasting between 50-100 days.
At very late times the light-curve is roughly consistent with nickel radioactive decay power \citep{Adams:2016MNRAS.460.1645A}. 

Optical spectra of ILRTs are also similar to those of Type II core-collapse SNe, showing strong hydrogen emission, however the features are not nearly as broad. 
While core-collapse SNe H$\alpha$ lines typically have velocity widths on the order of 10,000 km/s, the H$\alpha$  lines present in ILRT spectra are typically between 400-900 km/s \citep[e.g.,][]{Cai:2021A&A...654A.157C}. 
These lower velocities suggest lower explosion energies and masses than those of classical core-collapse SNe.
The early-time ($<$ 100 days) spectra of ILRTs are also characterized by strong calcium features including the Ca II IR triplet ($\lambda \lambda$ 8498, 8542, 8662) and [Ca II] doublet ($\lambda \lambda$ 7291, 7324) in emission, as well as the Ca II H and K lines ($\lambda \lambda$ 3968, 3934) in absorption \citep[e.g.,][]{Jencson:2019ApJ...880L..20J}. 
These features are superimposed on a very red continuum suggesting large quantities of dust and high extinction values. 
The overall set of absorption features (Ca II H and K, as well as the Na I D doublet ($\lambda \lambda$ 5889, 5895), and a strong O I blend near 7773 angstroms) resemble the spectrum of an F-type supergiant.
These features are also seen in some SN impostors/massive star eruptions and are expected for outbursts which form extended, optically thick winds \citep{Davidson:1987ApJ...317..760D}.

The physical processes which lead to ILRTs remain debated.
Since they are extremely self-enshrouded, it is not even known whether ILRTs are the result of terminal explosions or some kind of non-terminal stellar eruption. 
Three major hypotheses exist in the literature to explain the origins of ILRTs: common envelope mergers of high mass stars \citep[i.e. LRNe-like transients with more massive progenitor systems;][]{Kochanek:2014MNRAS.443.1319K}, massive eruptions of intermediate mass stars \citep[i.e. LBV-like giant eruptions of more intermediate mass stars;][]{Humphreys:2011ApJ...743..118H}, and electron capture supernovae of super-asymptotic giant branch (SAGB) stars \citep{Pumo:2009ApJ...705L.138P}.

Luminous red novae (LRNe) form another class of gap-transients. 
This class is typified by multi-peaked light curves and spectra which quickly evolve from those similar to CCSNe which show blue continuum and strong Balmer emission to much redder continuum and molecular lines \citep{Sparks:2008AJ....135..605S,Smith:2016MNRAS.458..950S, Blagorodnova:2017ApJ...834..107B, Karambelkar:2023ApJ...948..137K}. 
LRNe are typically explained as a post common-envelope phase of mass ejection in contact binaries which may be followed by a stellar merger \citep{Pastorello:2019A&A...630A..75P}.
In our own galaxy, V1309Sco showed long term periodic variability in its light curve due to the orbit of the binary prior to being observed as a LRNe helping to identify LRNe as the result of stellar merger events \citep{Tylenda:2011A&A...528A.114T}.
While early-time spectra of ILRTs and LRNe are similar, late-time spectroscopic observations show many differences, principally the appearance of molecular lines in LRNe spectra which are not seen in ILRT spectra. LRNe light curves are also distinct from ILRTs. LRNe light curves are fairly ``bumpy" with multiple peaks whereas ILRT light curves tend to have only a single, rather smooth, peak \citep{Karambelkar:2023ApJ...948..137K}. 
ILRTs are typically much more luminous than typical LRNe suggesting that if ILRTs are indeed stellar mergers they should be more massive merger events than typical LRNe.
It is possible that these higher masses also change the spectral features and light curve morphology and account for these differences between the ILRTs and the LRNe.

LBV giant eruptions, which are also called SN impostors because they appear like hydrogen rich interacting CCSNe \citep{Humphreys:1994PASP..106.1025H}, share many observational properties with ILRTs. Despite appearing very similar to terminal CCSNe, deep late time imaging of SN impostors reveal that the star itself has survived. During the eruption, very high mass loss rates (even exceeding 1 M$_{\odot}$ \citet{Humphreys:1999PASP..111.1124H}) can increase the luminosity of these objects by more than 3 magnitudes compared to their quiescent level. Like ILRTs, LBV eruptions show F-type supergiant-like absorption features, and have similar optical lightcurves, however they are much bluer and brighter.
For this reason it is expected that if ILRTs are also produced by stellar eruptions they should be the eruptions of more intermediate mass stars (8-15 M$_{\odot}$) whereas LBVs are all above 20-30 M$_{\odot}$ \citep{Smith:2004ApJ...615..475S}. 
LBV variability can be roughy divided into two classes: S Doradus-like visual brightening which has historically been thought to result not from a change in bolometric luminosity but from a shift toward the optical due to an expanding photosphere from a pulsation, possibly the result of super-Eddington luminosities in the sub-surface Fe-opacity bump, \citep{Humphreys:1999PASP..111.1124H, Smith:2014ARA&A..52..487S} and the Eta Carinae-like giant eruptions as described above in which the overall bolometric luminosity is changing significantly. 
ILRTs resemble the giant eruptions of LBVs in a dusty cocoon \citep{Bond:2009ApJ...695L.154B, Humphreys:2011ApJ...743..118H}.
The peak brightness of LBV giant eruptions is observed to range from -10 to -16 magnitudes in R band in a weakly bimodal distribution with a more populated brighter class peaking around -13 -- -15 mag and and a less populated fainter class peaking around -10 -- -11 mag \citep{Smith:2011MNRAS.415..773S}. Typical ILRTs peak between -11 -- -15 mag \citep{Cai:2021A&A...654A.157C} placing them on the lower-luminosity end of the brighter class of LBV giant eruptions.

Electron capture supernovae (ECSNe) are a theoretical class of faint SNe that come from the core-collapse of intermediate mass SAGB stars.
These stars, unlike their more massive counterparts, are unable to fuse up to iron in their cores \citep[e.g.,][]{Nomoto:1984ApJ...277..791N, Ritossa:1999ApJ...515..381R}.
Instead these stars finish burning with a degenerate oxygen-neon-magnesium core.
Electron capture onto the Ne and Mg nuclei removes the electrons whose degeneracy pressure was previously supporting the core resulting in core-collapse.
ECSNe are predicted to have lower energies and luminosities than typical, higher mass, iron core-collapse SNe, although there are many uncertainties associated with the theory \citep[e.g.,][]{Poelarends:2008ApJ...675..614P, Pumo:2009ApJ...705L.138P, Doherty:2015MNRAS.446.2599D, Doherty:2017PASA...34...56D}.
In part due to these uncertainties there has never been a confirmed detection of an ECSNe. 
ILRTs, which are less energetic than ordinary core-collapse SNe as well as having less massive progenitors are strong candidates for electron capture supernovae. 
Theoretical rate estimates of ECSNe are likewise consistent with the observed rate of ILRTs with the caveat that both are very uncertain \citep{Cai:2021A&A...654A.157C}.
In addition to ILRTs, low luminosity Type IIP events have also been proposed as ECSNe including SN 2018zd \citep{Hiramatsu:2021NatAs...5..903H}.


In this work we focus on AT 2019abn, originally designated ZTF 19aadyppr, which was discovered by the ZTF survey on MJD 22.6 January 2019 \citep{Jencson:2019ApJ...880L..20J}. 
It rose over 15 days to a peak r-band magnitude of -13, firmly in the gap between classical novae and supernovae, where it plateaued for a few months before fading below optical detection limits \citep{Jencson:2019ApJ...880L..20J, Williams:2020A&A...637A..20W}. 
AT 2019abn is located in a star forming spiral arm of the nearby galaxy M51a at only 8.6 Mpc \citep{McQuinn:2016ApJ...826...21M}.
The transient is still visible in the infrared and has photometry data points from both \textit{Spitzer} IRAC and the Wide-field Infrared Survey Explorer (WISE) as part of the NEOWISE mission.
JWST is uniquely suited to understanding these extremely red events and thus AT 2019abn was chosen as one of three unusual red transients to be observed by JWST as part of Cycle 1 GTO Program 1240 (PI: Ressler).

In this work, we publish the spectra taken by JWST and consider their implications for distinguishing the physical origins of ILRTs.
In section \ref{sec:Observations} we describe the data obtained with JWST and its reduction as well as photometry from NEOWISE, in section \ref{sec:Analysis} we analyze the PAH features present in the MIRI data as well as the late time MIR light curve, and in section \ref{sec:ILRT_origins} we discuss how our JWST spectra help to constrain the possible physical mechanisms suggested for ILRTs. Finally in section \ref{sec:Conclusions} we summarize the results of this work.

\section{Observations}
\label{sec:Observations}

\begin{figure*}[t!]
    \centering
    \includegraphics[scale=0.7]{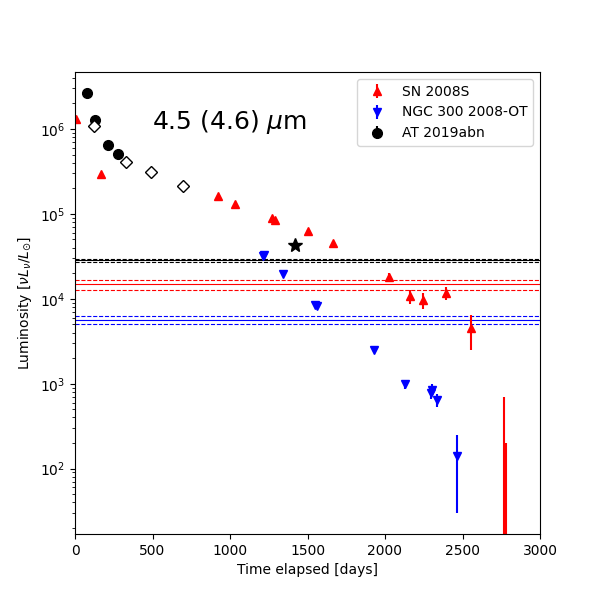}
    \caption{Photometry of the two prototypical ILRTs and AT 2019abn at 4.5 $\mu$m. The SN 2008S and NGC 300 2008-OT data points as well as the circular AT 2019abn points are \textit{Spitzer} IRAC Channel 2 photometry points. The unfilled diamond AT 2019abn points are from WISE Band 2 (4.6 $\mu$m) while the final star AT 2019abn point is a synthetic photometry point created by convolving the \textit{Spitzer} IRAC Channel 2 response with the JWST NIRSpec spectrum of AT 2019abn.}
    \label{fig:ILRT_photometry}
\end{figure*}

\subsection{JWST}
\label{sec:JWST_obs}
James Webb Space Telescope follow-up observations of AT 2019abn were taken as part of a Cycle 1 GTO program (PI Ressler, ID 1240) on UT 2022 December 13.5 (MJD=59926.5), 1,421 days after initial detection by ZTF on MJD 58505.6 and 1,396 days after the photometric peak of the optical transient on MJD 58530.6.
The JWST observations included fixed-slit NIRSpec \citep{Jakobsen:2022A&A...661A..80J} spectroscopy with the S2001A slit in the G395M/F290LP disperser-filter configuration which covers 2.87-5.10 microns with a spectral resolution R $\sim$ 1000, as well as fixed slit MIRI LRS \citep{Kendrew:2015PASP..127..623K} spectroscopy which covers 4.5-14 microns with a spectral resolution R $\sim$ 100.
Lower throughput resulting in lower signal to noise restricted the useful range of the MIRI LRS observations for this object to 4.6-12 $\mu$m. 
Raw (\texttt{uncal}) files were downloaded from the Mikulski Archive for Space Telescopes (MAST)\footnote{\url{https://mast.stsci.edu/portal/Mashup/Clients/Mast/Portal.html}} and reduced using a local installation of the standard JWST Science Calibration Pipeline \citep{bushouse_howard_2022_7229890}\footnote{\url{https://jwst-pipeline.readthedocs.io/en/stable/jwst/introduction.html}} version 1.12.5 under the the Calibration Reference Data System (CRDS; \citealp{greenfield2016}) context defined by \texttt{jwst\_1183.pmap}. In the \texttt{calwebb\_detector1} stage, we turned on \texttt{jump.expand\_large\_events} to reduce errors from cosmic rays on the detector.
The final 1D spectra were extracted using modified trace locations and background regions chosen by eye to reject emission from the the HII region surrounding the source using \texttt{extract\_1d} with a custom EXTRACT1D file. 
For the NIRSpec extraction the trace was extracted from pixels 22-26 along the dispersion axis (y-axis) with background subtraction from pixels 16-20 and 28-32.
For the MIRI LRS extraction the trace was extracted from pixels 29-33 along the dispersion axis (x-axis) with background subtraction from pixels 23-28 and 33-39. 
The combined NIRSpec and MIRI LRS spectra is shown in Figure \ref{fig:comb_spec}.
There is good agreement between the two instruments where they overlap around 5 microns helping to validate our choice of extraction region.

\subsection{Late-time WISE Photometry}
\label{sec:WISE_photometry}
The location of AT 2019abn is covered by the NEOWISE \citep{Mainzer:2011ApJ...731...53M} all-sky mid-IR survey in W1 (3.4 $\mu$m) and W2 (4.6 $\mu$m). In figure \ref{fig:ILRT_photometry} and table \ref{tab:photometry} we present W2 photometric data for the transient derived from a custom image subtraction pipeline \citep{De:2020PASP..132b5001D} based on the ZOGY algorithm \citep{Zackay:2016ApJ...830...27Z} as described in \citep[][supplementary information 3]{De:2023Natur.617...55D}. AT 2019abn is detected at 4.6 $\mu$m in 4 epochs before 1000 days post initial detection after which it falls below the detection limit of the unWISE time resolved co-added images \citep{Meisner:2023AJ....165...36M} from which the photometry has been obtained. 

\begin{deluxetable}{cc}
\tablecaption{Late-time MIR Photometry of AT 2019abn
    \label{tab:photometry}}
\tablehead{
  \colhead{\hspace{1.25cm}MJD}\hspace{1cm} & 
  \colhead{\hspace{1cm}Flux (10$^{-4}$ Jy)}\hspace{1.25cm}
  }
\startdata
58628.856 & 7.16 $\pm$ 0.13 \\
58835.384 & 2.68 $\pm$ 0.13\\
58995.736 & 2.08 $\pm$ 0.15\\
59200.886 & 1.42 $\pm$ 0.17\\
$\mathbf{59926.5}$ & $\mathbf{0.29 \pm 0.05}$\\
\enddata
\tablecomments{NEOWISE Band 2 (4.6 $\mu$m) photometry from image subtraction for AT 2019abn obtained as described in section \ref{sec:WISE_photometry}. AT 2019abn falls below the detection limits of NEOWISE in epochs after MJD 59200.886. The final bolded point is a synthetic photometry point obtained by convolving the JWST NIRSpec spectrum with the response function of \textit{Spitzer}/IRAC Channel 2 (4.5 $\mu$m). The error estimate on this synthetic photometry point should be taken as a lower limit.}
\end{deluxetable}

\section{Analysis}
\label{sec:Analysis}

\subsection{MIR PAH Features}
\label{sec:PAH_features}
Infrared spectroscopic observations have revealed many broad features which are ubiquitous in galactic and extragalactic sources. These features are (mostly) ascribed to large carbon molecules ($>$ 50 C atoms) called Polycyclic Aromatic Hydrocarbons (PAHs) \citep{Sellgren:1984ApJ...277..623S, Allamandola:1989ApJS...71..733A}.
The features are produced by a number of molecules which have similar carbon-carbon (C-C) or carbon-hydrogen (C-H) bonds. The C-C and C-H bonds are excited by FUV photons in various bending and stretching modes which when relaxed produce the broad features characteristic of PAH emission \citep[e.g.,][]{Draine:2001ApJ...551..807D}. 
PAHs are observed in many astronomical contexts including in HII regions, around post-AGB stars, in young stellar objects (YSOs), in reflection nebulae, and in galactic nuclei. PAH features are also observed in the general interstellar medium (ISM) of galaxies and are particularly observed in the IR cirrus and the surfaces of optically dark clouds \citep[e.g., ][and references therein]{Tielens:2008ARA&A..46..289T}. 
The diversity of PAH molecules is reflected in the diversity of the features they produce. Not only do the features produced depend on the structures of the individual PAH molecules but also on their density and the hardness of the input spectrum. They are commonly divided into multiple classes based on the position and relative strengths of the features observed as first done by \citet{Peeters:2002A&A...390.1089P}. Class A PAHs are observed where the ISM is illuminated by a star: in HII regions, reflection nebulae and the ISM, Class B PAHs are observed in CSM: in planetary nebulae, post-AGB sources, and in Herbig AeBe stars. Class C PAHs are the rarest class and are observed around unusual C-rich post-AGB sources or in carbon rich colliding wind Wolf-Rayet binary systems. Class A and class B PAH features are the most prevalent. Class C PAHs are much rarer and are typically observed only around carbon-rich sources. 

\begin{figure*}[t!]
    \centering
    \includegraphics[scale=0.7]{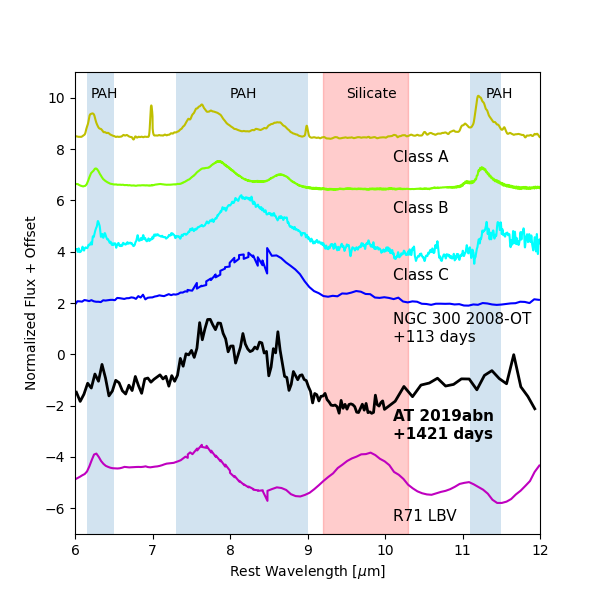}
    \caption{Mid-infrared spectra of AT 2019abn and NGC 200 2008-OT with MIRI LRS and \textit{Spitzer} IRS respectively as well as examples of various classes of PAH features. Note the similarities in the broad feature at 8.25 $\mu$m and also the differences in the 6.3 $\mu$m feature which is strongly present in the Class C spectrum, weaker in AT 2019abn, and entirely absent in NGC 300 2008-OT. We also show an example of a mixed chemistry LBV which shows Class B PAH features in addition to silicate features.}
    \label{fig:compare_to_class_C}
\end{figure*}

\begin{figure*}[t!]
    \centering
    \includegraphics[scale=0.7]{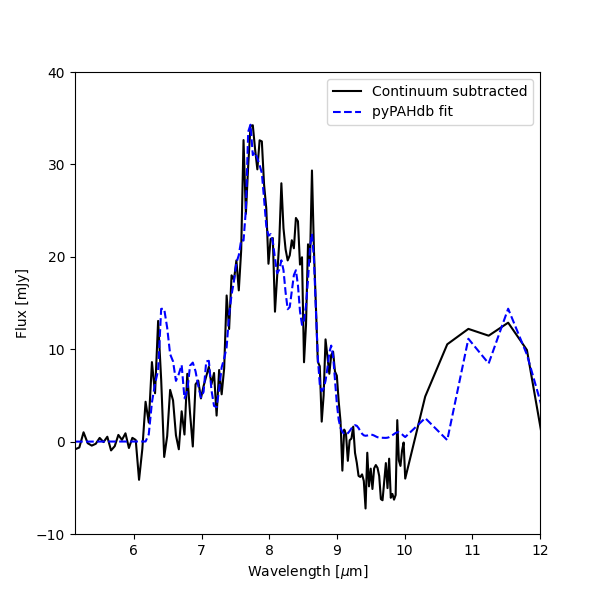}
    \caption{pyPAHdb fit of the continuum subtracted mid-IR spectrum of AT 2019abn. Note that the theoretical models predict a slightly stronger feature at 6.3 $\mu$m than is actually present. }
    \label{fig:pyPAHdb_fit}
\end{figure*}

The strongest features observed in the late-time JWST spectra of AT 2019abn are in the MIRI data.
These include a broad feature near 8.25 $\mu$m and a weaker feature near 6.3 $\mu$m. 
They seem to strongly resemble Class C PAH features. 
The feature at 8.25  $\mu$m is also seen in the only other ILRT for which a mid-IR spectra exists; NGC 300 2008-OT which has a \textit{Spitzer} IRS spectra showing PAH emission as reported in \citet{Prieto:2009ApJ...705.1425P}.
A comparison of the AT 2019abn MIRI spectrum with prototypical Class A, B, and C PAH sources and with the NGC 300 2008-OT spectrum is shown in figure \ref{fig:compare_to_class_C}.
We use the NASA AMES PAH database and python analysis package \citep{Boersma:2014ApJS..211....8B, Bauschlicher:2018ApJS..234...32B, Mattioda:2020ApJS..251...22M} \footnote{\url{https://github.com/PAHdb/pyPAHdb}} to model the PAH features observed in our spectrum after fitting and removing a continuum model.
A fit using the database models is shown in figure \ref{fig:pyPAHdb_fit} and reproduces the observed features in the spectrum.

\subsection{IR-Photometry of AT 2019abn}
\label{sec:ILRT_photometry}

Infrared light curves for SN 2008S, NGC 300 2008-OT, and AT 2019abn are shown in figure \ref{fig:ILRT_photometry}. 
Both SN 2008S and NGC 300 2008-OT are covered by \textit{Spitzer} and are shown to drop below their progenitor luminosities roughly 2000 days post-explosion suggesting that ILRTs are possibly terminal events, however without longer wavelength data it is not possible to rule out an additional cold dust component \citep{Adams:2016MNRAS.460.1645A}. 
AT 2019abn is covered early on by \textit{Spitzer} before it ended its mission and additional (4.6 $\mu$m) photometry points have been added from NEOWISE \citep{Mainzer:2011ApJ...731...53M} as described in section \ref{sec:WISE_photometry}.
By convolving the NIRSpec data with the \textit{Spitzer} IRAC Channel 2 response curve we have created an additional synthetic photometry point for AT 2019abn which still lies slightly above its progenitor luminosity.
The overall 4.5 $\mu$m light curves are roughly consistent between the ILRTs despite differences in progenitor and peak luminosities. 





\section{The Mysterious Origins Of ILRTs}
\label{sec:ILRT_origins}

There are three major hypotheses which exist in the literature to explain the physical origins of ILRTs: massive luminous blue variable (LBV) star-like massive eruptions of more intermediate mass stars \citep[SN impostor-like, e.g.,][]{Smith:2009ApJ...697L..49S, Andrews:2021ApJ...917...63A}, stellar mergers \citep[luminous red novae (LRNe)-like e.g.,][]{Kasliwal:2011ApJ...730..134K}, or electron capture supernovae \citep[ECSNe, e.g.,][]{Adams:2016MNRAS.460.1645A, Cai:2021A&A...654A.157C}.

\subsection{Higher mass stellar mergers?}
ILRTs represent a brighter population of gap transients than typical LRNe. If ILRTs are also produced by common-envelope ejections during a stellar merger their brighter peak luminosities should imply a more massive progenitor system. 
There is a theoretical \citep{Ivanova:2013Sci...339..433I} and empirically observed \citep{Kochanek:2014MNRAS.443.1319K} relationship between the peak luminosity of LRNe and the mass of the progenitor system. 
While the empirical relationship derived in \citet{Kochanek:2014MNRAS.443.1319K} is based on galactic mergers (which are a lower luminosity population as compared to the observed extragalactic events) two additional extragalactic LRNe extend this relationship to higher progenitor masses and peak luminosities: NGC 4490 OT2011-1 which peaked at -15 mag has an estimated progenitor mass of 20-30 M$_{\odot}$ \citep{Smith:2016MNRAS.458..950S} and M101 OT2015-1 which peaked at M$_{r} \leq$ -12.4 mag (lower limit due to gap in observations) has an estimated progenitor mass of $\approx$ 18 M$_{\odot}$ \citep{Blagorodnova:2017ApJ...834..107B}. AT 2019abn has a peak r band magnitude of -13 mag.
If the $L_{peak} \propto M_{prog}^{2-3}$ relationship from \citet{Kochanek:2014MNRAS.443.1319K} holds this would imply a 15-20 M$_{\odot}$  progenitor for AT 2019abn.

A 15-20 M$_{\odot}$ progenitor is in disagreement with the estimated progenitor mass for AT 2019abn derived from the pre-outburst \textit{Spitzer} images in \citet{Jencson:2019ApJ...880L..20J} of 8-15 M$_{\odot}$.
The Class C PAH features observed in our MIRI spectrum are also inconsistent with a more massive progenitor which are expected to be more oxygen rich even in the case where the \textit{Spitzer} observations underestimate the mass of the progenitor because of dust obscuration.

Overall we disfavor a LRNe/common-envelope ejection interpretation for AT 2019abn.
The spectral features observed in the optical spectra do not match with other LRNe, and its optical light curve does not show the multipeaked structure of observed LRNe. The peak luminosity of AT 2019abn in the LRNe case would also suggest a more massive progenitor which is seemingly ruled out by the \textit{Spitzer} observations of the progenitor and the observation of Class C PAH features in our JWST MIRI spectrum which are unexpected for a more massive progenitor as they are typically oxygen rich.

\subsection{Intermediate mass stellar eruptions or dusty LBV eruptions?}

We use LBV-like eruption after \citet{Smith:2011MNRAS.415..773S} to refer to a broad class of non-terminal eruptions of massive stars.
The optical spectra of AT 2019abn, as well as other ILRTs, resemble the optical spectra of other LBV-like giant eruptions \citep{Jencson:2019ApJ...880L..20J}.
From the pre-outburst imaging and the red color of AT 2019abn, as with other ILRTs, we know that the progenitor of AT 2019abn was heavily dust obscured prior to the observation of this transient.
It is known that LBV-like eruptions can be very non-spherical from observations of the Homunculus nebula produced by the Great Eruption of Eta Carinae \citep[e.g.,][]{Steffen:2014MNRAS.442.3316S}.
The diversity of observed expansion velocities in extra-galactic LBV-like eruptions (100-1000 km/s) can also be interpreted in the context of non-spherical eruptions observed at a variety of viewing angles \citep{Smith:2011MNRAS.415..773S}. 
It is possible to envision a scenario where the progenitor of AT 2019abn experienced previous episodes of mass loss which resulted in an anisotropic dusty torus which scatters light away from our line of sight and out along the poles.
In this scenario both the progenitor luminosity (and therefore mass) as well as the peak brightness of the transient itself would be underestimated.
Correcting for this effect would possibly bring the observed properties of AT 2019abn more into line with the observed properties of other LBV-like giant eruptions depending on the efficiency of the dust torus's redirection of starlight. 
In this view transients observed as ILRTs and transients observed as LBV-like eruptions are actually members of the same class except that ILRTs are observed at a different viewing angle and/or with a more recent history of mass loss \citep[e.g., the case of AT 2019krl][]{Andrews:2021ApJ...917...63A}.  

The giant eruptions of massive stars are themselves not well understood. In all cases the Eddington ratios of these eruptions are greater than unity, resulting in high rates of mass loss. For the less extreme cases (Eddington ratios of order a few) the eruptions may be explained by super-Eddington winds however it is unclear where the energy needed to drive the luminosity super-Eddington originates \citep{Smith:2014ARA&A..52..487S}. For more extreme eruptions (including the Great Eruption of Eta Car) an explosive origin is more likely and more in line with observation. A variety of explanations have been explored to understand how the energy needed to drive the explosion (or Super-Eddington wind) is provided including shell burning instabilities \citep{Dessart:2010MNRAS.405.2113D, Smith:2014ApJ...785...82S}, mass loss driven by gravity waves \citep{Meakin:2007ApJ...667..448M, Quataert:2012MNRAS.423L..92Q, Shiode:2014ApJ...780...96S}, stellar collisions or mergers \citep{Podsiadlowski:2010MNRAS.406..840P, Smith:2011MNRAS.415..773S, Smith:2014ApJ...785...82S, Hirai:2021MNRAS.503.4276H}, or extreme mass transfer events \citep{Smith:2014ARA&A..52..487S}.

There are two scenarios for which we can envision an LBV-like eruption origin for AT 2019abn. 
In the first scenario the lower observed luminosities of both transient and progenitor as compared to other known LBVs are intrinsic to a lower mass progenitor. 
In this case it is difficult to explain why a lower mass progenitor might experience the same kinds of eruptive mass loss observed in more massive stars as these lower mass progenitors are not expected to evolve close to their Eddington limits the way the more massive progenitors do.
In the second scanario the observed luminosities are artificially lowered as the result of a dusty torus which scatters light away from our line of sight and the true luminosities of both transient and progenitor are more in line with typical LBV-like eruptions.
It is unclear how a more massive progenitor would form the carbonaceous dust necessary to explain the observation of Class C PAH features in our JWST MIRI spectrum.

One caveat to this interpretation is that there do exist mixed chemistry LBVs. 
There are a small number of LBVs with PAH features observed alongside silicate features from more oxygen rich dust. It is unknown what leads to this mixed chemistry especially as theoretically massive stars like LBVs should have circumstellar environments with C/O less than 1 \citep{Sarangi:2015A&A...575A..95S, Sarangi:2018SSRv..214...63S}. Those LBVs which show PAH features include R71 in the Large Magellanic Cloud (LMC) for which a MIR spectrum is shown in figure \ref{fig:compare_to_class_C} \citep{Guha:2014A&A...569A..80G}, AG Car \citep{Voors:2000A&A...356..501V}, and HD 168625 \citep{Skinner:1997ASPC..120..322S, Umana:2010ApJ...718.1036U}. For those LBVs which do show PAH features they are exclusively Class B and are generally weaker than the strong silicate features present. This distinguishes them from the Class C PAH features observed in our spectrum of AT 2019abn which also does not show silicate emission. For R71 in the LMC the mass of PAHs is estimated to be less than 5 $\%$ of the total dust mass \citep{Guha:2014A&A...569A..80G}. 

Overall we consider a LBV-like eruption scenario for AT 2019abn to be less likely. The ultimate test would be to look for a surviving dust-obscured luminous star at longer wavelengths with JWST at even later times.

\subsection{Electron capture supernovae?}
ILRTs have been put forth as candidate ECSNe because they exhibit many of the observational properties theory predicts.
Their progenitors, when detected, are consistent with SAGB stars \citep{Prieto:2008ApJ...681L...9P, Bond:2009ApJ...695L.154B, Botticella:2009MNRAS.398.1041B, Thompson:2009ApJ...705.1364T, Jencson:2019ApJ...880L..20J}.
AGB stars undergoing thermal pulsation are excellent producers of dust.
Modeling suggests that SAGB stars ought to also undergo thermal pulsation which would lead to mass loss \citep{Jones:2013ApJ...772..150J} which is consistent with the dust obscured progenitors observed for ILRTs prior to a terminal explosion. 

SAGB progenitors which are thought to be between 8 and 12 M$_{\odot}$ are more massive than the typical carbon-rich dust producing AGB stars which are expected to be between 3 and 5 M$_{\odot}$ \citep{Ventura:2012MNRAS.420.1442V, Ventura:2012MNRAS.424.2345V}. There is disagreement in the literature about the expected C/O ratio in the envelopes of SAGB stars. The relative rates of carbon and oxygen burning depend on the temperature of Hot Bottom Burning (HBB) during the third dredge-up (TDU). 
The most massive SAGB stars can have HBB at higher temperatures leading to efficient oxygen burning over carbon burning and higher C/O ratios.
Therefore unlike in typical red supergiants the CSM of SAGB stars may be expected to be O-poor \citep{Jones:2013ApJ...772..150J, Doherty:2017PASA...34...56D, Hiramatsu:2021NatAs...5..903H}.
We note that the relative rate of carbon and oxygen burning in HBB is also sensitive to metallicity with metal poor stars becoming carbon-rich at lower masses than those at solar metallicity \citep{Doherty:2014MNRAS.437..195D} with the most metal rich stars perhaps never becoming carbon-rich at all. 

Overall our observations of Class C PAH features in AT 2019abn are most consistent with the ECSNe hypothesis. Carbonaceous dust grains are allowed in the SAGB progenitor scenario in the case of high temperature HBB where oxygen is very efficiently burned leading to a C/O ratio greater than 1.
A terminal ECSNe explosion is also consistent with the observed fading of SN 2008S and NGC 200 2008-OT \citep[Figure 2 and ][]{Adams:2016MNRAS.460.1645A}, with the caveat that a cool dust shell or torus could be reprocessing/scattering away light and concealing an extant luminous star.

We note that it has been suggested that the progenitor of AT 2019abn is also consistent with a $\sim$ 15 M$_{\odot}$ dust-obscured RSG \citep{Jencson:2019ApJ...880L..20J}. However the lower explosion energies and luminosities of AT 2019abn and other ILRTs as compared to typical core-collapse SNe are difficult to reproduce in theoretical models of RSG explosions \citep{Moriya:2024arXiv240712284M}.
Furthermore the vast majority of dusty RSGs are observed to have oxygen rich silicate dust and no PAH features, which is inconsistent with our observations of class C PAH features in AT 2019abn \citep{Verhoelst:2009A&A...498..127V, Wang:2021ApJ...912..112W}.  
There are an extremely small number of RSGs which do show PAH features, including U Lacertae and RS Persei but these few outliers are unusual in other ways; RS Persei is possibly a lower mass evolved AGB star \citep{Yoon:2014ApJS..211...15Y} and U Lacertae has an unresolved, hot, possibly post-AGB, companion \citep{Burki:1983A&A...124..256B}.
For these reasons we consider a RSG explosion scenario much less likely than a SAGB ECSNe scenario.

\section{Conclusions}
\label{sec:Conclusions}

In this work, we have presented late-time JWST NIR/MIR observations of the intermediate-luminosity red transient (ILRT) AT 2019abn. Our observations show class C PAH features typical of carbon-rich post-AGB sources suggesting that the dust around AT 2019abn is composed primarily of carbonaceous grains. These rare class C PAH features are also observed in the only other ILRT for which a MIR spectrum exists, NGC 300 2008-OT, which was observed with \textit{Spitzer} IRS. We also present late-time photometry of AT 2019abn, obtained with WISE and JWST. While AT 2019abn continues to fade at a rate similar to SN 2008S, it is not yet below the progenitor luminosity level. 

Three major hypotheses to explain the origins of ILRTs exist in the literature: non-terminal stellar eruptions, stellar merger events, and terminal ECSNe. In light of our new JWST observations we conclude that:
\begin{enumerate}
    \item  Since we observe Class C PAH features and no silicate features, ILRTs have a CSM which is composed of carbonaceous dust grains with a C/O ratio greater than 1.
    \item Carbonaceous dust grains and a high C/O ratio suggest that the progenitors of ILRTs are consistent with SAGB stars with high temperature HBB. Furthermore, the observed Class C PAH features are inconsistent with massive stars or typical red supergiants. Overall our JWST observations are consistent with the ECSNe scenario. 
    \item Continued monitoring of the late-time decline of AT2019abn at longer wavelengths and later time with JWST is necessary to test whether or not the explosion was indeed terminal. 
\end{enumerate}

The presence of Class C PAH features in the spectra of AT 2019abn strongly suggests that the dust produced by ILRTs (or their progenitors) is carbon-rich.
Recent studies with JWST have showed that dust is formed in significant quantities even at very early times (within 600 million years of the Big Bang), including dust particles with carbon-rich compositions that may be attributed to dust-forming carbon-rich Wolf-Rayet stars \citep{Lau:2022NatAs...6.1308L} or supernova ejecta \citep{Witstok:2023Natur.621..267W}.
However supernovae, which have previously been invoked to explain the bulk of early universe dust formation, come from massive progenitors and are expected to form mostly oxygen rich silicate dust \citep[e.g.,][]{Gall:2011A&ARv..19...43G, Sarangi:2015A&A...575A..95S, Sarangi:2018SSRv..214...63S}.
It is well-know that in the local universe carbon-rich dust can be formed in significant amounts by low mass AGB stars \citep{Ventura:2012MNRAS.420.1442V, Ventura:2012MNRAS.424.2345V}, however these low mass stars are too long-lived to explain this early dust formation. 
Perhaps ILRTs, which may have more massive and thus shorter lived progenitors compared to AGBs, can help to explain the presence of carbon-rich dust at such early times. Progress requires a larger sample of ILRTs that are well-studied in the infrared. 



Upcoming and future infrared surveys, such as WINTER, DREAMS, PRIME, Cryoscope and the Nancy Grace Roman Space Telescope \citep{Akeson:2019arXiv190205569A}, will overcome the limitations of current optical surveys in uncovering large, unbiased samples of dust-obscured transients such as ILRTs. ILRTs will be readily discovered as IR transients in alerts produced for Roman with the RAPID\footnote{The Roman Alerts Promptly from Image Differencing (RAPID) Project Infrastructure Team: \url{https://www.ipac.caltech.edu/project/rapid}} pipeline, and identified by their distinctive red optical colors provided by Rubin  \citep{Ivezic:2019ApJ...873..111I}. Spectroscopic follow-up of ILRTs with JWST will quantify their contribution to the cosmic dust budget.

\section{Acknowledgements}
\label{sec:Acknowledgements}
We thank our JWST program coordinator Shelly Meyett for facilitating our Target of Opportunity (ToO) Activation Request. 
We are grateful to NIRSpec Instrument Scientist Tony Keyes for his critical input on target acquisition for our NIRSpec observations.
The authors thank Carolyn Doherty for very helpful discussions regarding the chemical composition of SAGB stars. We also thank Arkaprabha Sarangi and Tea Temim for discussions on dust formation in supernovae.  We are also grateful to Emma Beasor for discussions related to red transients and JWST. 
We acknowledge support from grants JWST-ERS-01349.002-A and JWST-GO-01863.002-A. 
MMK and JJ acknowledge NASA support for the RAPID project infrastructure team under award 80NSSC24M0020 (program NNH22ZDA001N-ROMAN). 

\facilities{JWST (MIRI LRS, NIRSpec), NEOWISE}

\software{JWST Science Calibration Pipeline \citep{bushouse_howard_2022_7229890}, Jdaviz \citep{jdadf_developers_2023_7971665}, Astropy \citep{Astropy:2013, Astropy:2018}, Matplotlib \citep{Hunter:2007}, NumPy \citep{Walt:2011}, SciPy \citep{Virtanen:2019}, pyPAHdb \citep{Boersma:2014ApJS..211....8B, Bauschlicher:2018ApJS..234...32B, Mattioda:2020ApJS..251...22M}}

\bibliographystyle{aasjournal}
\bibliography{main.bib}

\end{document}